\documentclass[reprint,aps,prc,showpacs,preprintnumbers,superscriptaddress,amsmath,amssymb,twocolumn,10pt,floatfix]{revtex4-2}

\usepackage{amsmath,amssymb,latexsym,amsthm}
\usepackage{bm}
\usepackage{CJKutf8}
\usepackage{graphicx}
\usepackage{xcolor}
\usepackage{ulem}

\begin{document}

\title{
  Axially Deformed Proton-Neutron Relativistic Quasiparticle Finite Amplitude Method for Charge-Exchange Transitions
}

\author{Chen Chen (\begin{CJK}{UTF8}{gbsn}陈晨\end{CJK})}
\affiliation{Frontiers Science Center for Rare isotope, Lanzhou University, Lanzhou 730000, China}
\affiliation{School of Nuclear Science and Technology, Lanzhou University, Lanzhou 730000, China}

\author{Zhengzheng Li (\begin{CJK}{UTF8}{gbsn}李征征\end{CJK})}
\affiliation{State Key Laboratory of Nuclear Physics and Technology, School of Physics, Peking University, Beijing 100871, China}

\author{Yinu Zhang (\begin{CJK}{UTF8}{gbsn}张一怒\end{CJK})}
\affiliation{Sino-French Institute of Nuclear Engineering and Technology, Sun Yat-sen University, Zhuhai 519082, China}

\author{Yifei Niu (\begin{CJK}{UTF8}{gbsn}牛一斐\end{CJK})}\email[]{niuyf@lzu.edu.cn}
\affiliation{Frontiers Science Center for Rare isotope, Lanzhou University, Lanzhou 730000, China}
\affiliation{School of Nuclear Science and Technology, Lanzhou University, Lanzhou 730000, China}

\date{\today}

\begin{abstract}
  The quasiparticle finite amplitude method (QFAM)
  is extended to describe charge-exchange transitions
  based on the relativistic Hartree-Bogoliubov model,
  adopting the point-coupling energy density functional DD-PC1
  and a finite-range separable pairing force.
  After validation through comparison with relativistic quasiparticle random-phase approximation (QRPA) results in spherical nuclei,
  the deformation effects on isobaric analog resonances (IAR) and Gamow-Teller (GT) transitions
  in Zn isotopes are investigated.
  The GT strength exhibits significant fragmentation in deformed nuclei.
  The analysis of summed strengths and centroid energies in GT resonance region between the $K=0$ and $K=1$ components reveals
  that prolate configurations exhibit  stronger $K=1$ strength and lower $K=1$ centroid energy, while oblate shapes show an opposite behavior, with stronger $K=0$ strength and lower $K=0$ energy.
  The effects of isoscalar pairing on GT strength distributions for different shape configurations are also examined.
\end{abstract}

\maketitle

\section{Introduction\label{sec:intro}}

Nuclear charge-exchange transitions involve a neutron becoming a proton or vice versa in the nucleus,
whose typical modes include isobaric analog resonance (IAR) and Gamow-Teller (GT) transitions \cite{FOsterfeld_1992_RMP}.
The study of charge-exchange transitions is not only a central topic in nuclear physics \cite{YFujita_2011_PPNP},
but is also crucial for determining $\beta$-decay rates, serving as key inputs in rapid neutron-capture process simulations
\cite{TKajino_2019_PPNP, ZMNiu_2019_PRC, YWHao_2023_PLB},
and predicting nuclear matrix elements of neutrinoless double $\beta$-decay \cite{JMYao_2022_PPNP},
a key factor in  determining the neutrino mass hierarchy.

Despite significant experimental efforts,
obtaining comprehensive experimental data across the entire nuclide chart remains challenging.
Consequently, theoretical predictions of charge-exchange excitations are crucial for addressing applications in nuclear astrophysics and guiding future experiments.
Various theoretical models, such as ab-initio approaches \cite{HHerget_2020_FP, PGysbers_2019_NP, SPastore_2018_PRC},
shell model \cite{ECaurier_2005_RMP, QZhi_2013_PRC, GMartinezPinedo_2022_physics},
and random-phase approximation (RPA) \cite{JSuhonen_2007_FNTN, Colo_2020_Handbook}, have been developed to describe
charge-exchange excitations. Among these models, RPA offers the most computationally efficient framework for global descriptions.

Within the RPA framework, charge-exchange excitations are described as coherent superpositions of proton-neutron particle-hole (p-h) excitations,
commonly referred to as proton-neutron RPA (pnRPA) \cite{JSuhonen_2007_FNTN}.
For superfluid nuclear systems, this formalism naturally extends to two-quasiparticle (2qp) excitations,
giving rise to the proton-neutron quasiparticle RPA (pnQRPA) approach \cite{Colo_2020_Handbook}.
When combined with density functional theory (DFT),
the pnQRPA enables a self-consistent description of both ground-state properties and excitation spectra with the same energy density functional and pairing interaction.
Self-consistent pnQRPA approaches have been successfully implemented based on various DFTs,
including relativistic DFT \cite{Paar_2004_pnRQRPA, DVretenar_2008_JPG} and Skyrme DFT \cite{JEngel_1999_PRC, SFracasso_2005_PRC}.
The self-consistent pnQRPA method has demonstrated its success
in describing charge-exchange nuclear transitions \cite{MBender_2002_PRC, SFracasso_2007_PRC, NPaar_2005_EPJA, ZMNiu_2017_PRC},
$\beta$-decay rates \cite{TNiksic_2005_PRC, Marketin_2016_PRC}
and double $\beta$-decay nuclear matrix elements \cite{NPopara_2022_PRC, WLLv_2022_PRC, WLLv_2023_PRC}.

Above pnQRPA studies are typically restricted to the spherical symmetry assumption,
despite the fact that nuclear deformation is a common feature across the nuclide chart \cite{KYZhang_2022_ADNDT, PGuo_2024_ADNDT}.
Recent developments incorporating deformation degrees of freedom in pnQRPA calculations based on Skyrme \cite{KYoshida_2013_PTEP} and Gogny \cite{Martini_2014_NDS} DFTs
have demonstrated that nuclear deformation can significantly improve the description of charge-exchange transitions \cite{Mustonen_2013_PRC,Martini_2014_PRC}
and consequently enhance the accuracy of $\beta$-decay half-life predictions~\cite{Sarriguren_1999_NPA,  Yoshida_2023_PRC},
which in turn helps to further constrain ground-state deformations~\cite{Nacher_2004_PRL}
and potentially reduces uncertainties in nucleosynthesis simulations.
However, incorporating deformation into traditional pnQRPA calculations presents significant computational challenges,
as the single-particle level splitting in deformed nuclei complicates both the construction of QRPA matrix elements
and the diagonalization of large matrices, particularly in heavy nuclei.

To circumvent the computationally expensive problem in traditional QRPA calculations,
the finite amplitude method (FAM) \cite{Nakatsukasa_2007_FAM, HZLiang_2013_PRC} and its extension to superfluid nuclei,
quasiparticle FAM (QFAM) \cite{Avogadro_2011_QFAM, Niksic_2013_QFAM},
have been developed.
The (Q)FAM approach offers a more efficient solution to the (Q)RPA equations
by employing an iterative scheme within the linear-response framework,
thus avoiding both matrix diagonalization and the explicit evaluation of two-body matrix elements.
Initially, the QFAM method was successfully applied to non-charge-exchange transitions
within both non-relativistic Skyrme DFT \cite{Avogadro_2011_QFAM,MStoitsov_2011_PRC,JCPei_2014_PRC} and relativistic DFT frameworks \cite{Niksic_2013_QFAM, XWSun_2017_QFAM, Bjelcic_2020_QFAM, XWSun_2022_QFAM},
achieving its success in describing nuclear multipole resonances
\cite{TInakura_2009_PRC, TInakura_2011_PRC, Kortelainen_2015_PRC, TOishi_2016_PRC, CChen_2025_arXiv}.
Subsequently, this method was extended to charge-exchange transitions based on non-relativistic Skyrme DFT \cite{Mustonen_2014_pnQFAM, Shafer_2016_PRC},
and applied to large-scale calculations of $\beta$-decay rates~\cite{Mustonen_2016_PRC, EMNey_2020_PRC}
as well as studies of two-neutrino double-$\beta$ decay~\cite{Hinohara_2022_PRC}.

Compared to non-relativistic DFTs, such as those of the Skyrme type,
relativistic DFTs offer certain advantages that naturally incorporate the spin-orbit interaction
and inherently satisfy Lorentz invariance, thereby reducing the number of adjustable parameters~\cite{JMeng_2006_RCHB,JMeng_2016_book}.
Nevertheless, the development of relativistic QFAM for charge-exchange transitions remains absent.
Therefore, the present work aims to establish a relativistic proton-neutron QFAM (pnQFAM) approach,
providing an efficient framework for describing charge-exchange transitions in deformed nuclei
within the relativistic formalism,
and to apply this method to the study of Fermi and GT transitions in such systems.

Previous studies incorporating the deformation degree of freedom in charge-exchange transitions
have shown that deformation significantly fragments the GT transition strength due to the splitting between two modes with different projected total angular momentum $K$
~\cite{PSarruguren_2006_IJMPE, MMartini_2014_PRC,EJHa_2024_PRC}.
On the other hand, in the study of isovector dipole giant resonance, a typical non-charge-exchange mode,  the systematic trends in the summed strengths and centroid energies of the $K$ splitting in prolate and oblate nuclei have been clearly investigated~\cite{Danos_1958_NP, YXu_2021_PRC, MMartini_2016_PRC, YTian_2019_CPC, CChen_2025_arXiv}.
However, such kinds of systematics and its behind mechanisms have not been investigated in detail for GT transitions.
By taking advantage of the relativistic pnQFAM approach developed in this work,
we will  not only systematically investigate the strength distributions but also try to discover the systematics of the $K$ splitting exhibited in GT resonances and reveal the mechanisms behind such behaviors.

This work is organized as follows:
In Sec.~\ref{sec:theor}, we introduce the formalism of the relativistic pnQFAM approach for charge-exchange transitions,
detailing both particle-hole and particle-particle interactions.
In Sec.~\ref{sec:num}, we provide the numerical details and compare the results with the relativistic QRPA model.
Finally, in Sec.~\ref{sec:res}, we present the results for both Fermi and GT strengths in Zn isotopes,
followed by a discussion of the deformation effects on strength distributions, as well as summed strengths 
and centroid energies of different $K$ modes in the GT resonance (GTR) region. 
The effect of isoscalar pairing is also investigated for different shape configurations.

\section{Theoretical Framework\label{sec:theor}}
In this section, the formalism of the pnQFAM based on the axially deformed relativistic Hartree-Bogoliubov (RHB) theory is introduced.
The axially deformed RHB equation is formulated and solved in the axially deformed harmonic oscillator (HO) basis.
The RHB model provides a unified description of nuclear particle-hole (ph)
and particle-particle (pp) correlations on the mean-field level
by the introduction of a unitary Bogoliubov transformation to Bogoliubov quasiparticles.
The RHB Hamiltonian $H$ is obtained from the variation of an energy density functional $E[R]$
with respect to the generalized density matrix $R$,
which includes the density matrix and the pairing tensor.
As a result, two kinds of average potentials are incorporated into the RHB equation,
namely, the self-consistent nuclear mean field that encloses all the long-range ph correlations,
and a pairing field that sums up the pp correlations.
Details can be found in Ref.~\cite{Niksic_2014_DIRHB}.
In the following, the pnQFAM equations are introduced in detail,
together with the corresponding induced single-particle Hamiltonian and induced pairing field
formulated within the density-dependent point-coupling relativistic framework plus a finite-range separable pairing force.

\subsection{Proton-neutron quasiparticle finite amplitude method for charge-exchange transitions}
The derivation of QFAM method has been introduced in Refs.~\cite{Avogadro_2011_QFAM,Niksic_2013_QFAM,Mustonen_2014_pnQFAM,XWSun_2017_QFAM,Bjelcic_2020_QFAM}.
The starting point is the time-dependent RHB equation,
\begin{equation}\label{eq:tdfhb}
  \mathrm{i}\hbar \dot{R}(t) = [H\left[R(t)\right]+F(t),R(t)].
\end{equation}
The above equation describes the response of generalized density matrix $R(t)$ under an external field $\hat{F}(t)$.
$\hat{F}(t)$ can be expressed as
\begin{equation}\label{eq:externalfield}
  \hat{F}(t)
  = \hat{F} \mathrm{e}^{-\mathrm{i} \omega t}
  + \hat{F}^{\dagger} \mathrm{e}^{\mathrm{i} \omega t},
\end{equation}
with $\omega$ the oscillation frequency of the external field.

Assuming the external field $\hat{F}$ is weak,
which introduces only oscillations around the stationary density,
\begin{equation}
  R(t)= R_0 + \delta R(\omega) \mathrm{e}^{-\mathrm{i} \omega t}+\delta R^{\dagger} (\omega) \mathrm{e}^{\mathrm{i} \omega t},
\end{equation}
the RHB Hamiltonian evolves as follows,
\begin{equation}
  H(t)=H_{0}+\delta H(\omega) \mathrm{e}^{-\mathrm{i} \omega t}+\delta H^{\dagger} (\omega) \mathrm{e}^{\mathrm{i} \omega t}.
\end{equation}
Keeping the linear terms in the equation of motion of Eq. \eqref{eq:tdfhb},
and applying the Fourier transformation,
one can obtain the linear response equation in the frequency domain,
\begin{equation}\label{eq:lre}
  \omega \delta R(\omega)
  = \left[H_0, \delta R(\omega)\right]
  + \left[\delta H(\omega)+F, R_0\right] .
\end{equation}
In the following, it is more straightforward to derive the QFAM equation
in the stationary Bogoliubov quasiparticle representation, i.e.,
\begin{equation}\label{eq:qplre}
  \omega \delta\mathcal{R}(\omega)
  = \left[\mathcal{H}_0, \delta\mathcal{R}(\omega)\right]
  + \left[\delta\mathcal{H}(\omega)+\mathcal{F}, \mathcal{R}_0\right] .
\end{equation}
Here, the convention that calligraphic symbols correspond to matrices in the Bogoliubov quasiparticle basis is adopted.
In this representation, the Hamiltonian $\mathcal{H}_0$ and
the generalized density matrix $\mathcal{R}_0$ are diagonal,
\begin{equation}
  \mathcal{H}_0=\left(\begin{matrix} \mathcal{E} & 0 \\ 0 & -\mathcal{E} \end{matrix}\right), \quad
  \mathcal{R}_0=\left(\begin{matrix} 0 & 0 \\ 0 & 1 \end{matrix}\right).
\end{equation}
Because $\mathcal{R}(t)$ is a projector, i.e., $\mathcal{R}(t)^2 = \mathcal{R}(t)$,
only the off-diagonal blocks of $\delta\mathcal{R}(t)$ can be nonzero,
\begin{equation}
  \delta\mathcal{R}(\omega)
  \equiv \left(\begin{matrix} 0 & \mathcal{X}(\omega) \\ -\mathcal{Y}(\omega) & 0 \end{matrix}\right) .
\end{equation}
By substituting $\mathcal{R}_0$, $\delta\mathcal{R}$, and $\mathcal{H}_0$
into the linear response equation~\eqref{eq:qplre},
one can obtain the QFAM equations,
\begin{equation}
  \begin{aligned}
    \left(\mathcal{E}_{\mu}+\mathcal{E}_{\mu^\prime}-\omega\right)\mathcal{X}_{\mu\mu^\prime}(\omega) + \delta\mathcal{H}_{\mu\mu^\prime}^{20}(\omega)=-\mathcal{F}_{\mu\mu^\prime}^{20}, \\
    \left(\mathcal{E}_{\mu}+\mathcal{E}_{\mu^\prime}+\omega\right)\mathcal{Y}_{\mu\mu^\prime}(\omega) + \delta\mathcal{H}_{\mu\mu^\prime}^{02}(\omega)=-\mathcal{F}_{\mu\mu^\prime}^{02},
  \end{aligned}
\end{equation}
where $\mu$ and $\mu^\prime$ run over both the neutron Bogoliubov quasiparticle state $\nu$
and proton Bogoliubov quasiparticle states $\pi$.
The induced Hamiltonian $\delta\mathcal{H}(\omega)$
depends on the induced generalized density matrix $\delta\mathcal{R}(\omega)$.
In practice, these equations are solved iteratively and involve only the first derivatives of $E[R]$ with respect to $R$.
Such an approach avoids tedious calculations of two-body matrix elements and matrix diagonalization,
offering a more efficient approach for studies of deformed nuclei.

In the following, we will concentrate on the charge-exchange transitions.
For the one-body  charge-exchange transition operator in the $\beta^-$ direction,
the external field $\hat{F}$ can be written in a single-particle basis as,
\begin{equation}
  \hat{F}=\sum_{pn} f_{pn} c^{\dagger}_{p} c_{n}.
\end{equation}
$f_{pn}$ is the single-particle transition matrix element from a neutron state $n$ to a proton state $p$.
In the Bogoliubov quasiparticle basis, $\hat{F}$ can be expressed as,
\begin{equation}
  \hat{F}=\sum_{\pi\nu} \left[\mathcal{F}^{20}_{\pi\nu} \beta^{\dagger}_{\pi} \beta^{\dagger}_{\nu}
    + \mathcal{F}^{02}_{\pi\nu} \beta_{\nu} \beta_{\pi}
    + \cdots\right].
\end{equation}
Here, $\beta^{\dagger}_{\pi} (\beta^{\dagger}_{\nu})$
and $\beta_{\pi} (\beta_{\nu})$
correspond to the creation and annihilation of proton (neutron) Bogoliubov quasiparticles, respectively, with the corresponding matrix elements,
\begin{equation}
  \mathcal{F}^{20}_{\pi\nu} =   \sum_{pn} U^\ast_{p \pi} f_{pn} V^\ast_{n \nu},\quad
  \mathcal{F}^{02}_{\pi\nu} = - \sum_{pn} V_{p \pi} f_{pn} U_{n \nu},
\end{equation}
where $U$ and $V$ are the Bogoliubov transformation matrices obtained from the RHB model.

The QFAM equations in the charge-exchange case become the proton-neutron QFAM (pnQFAM) equations,
\begin{equation}\label{eq:pnqfam}
  \begin{aligned}
    \left(\mathcal{E}_{\pi}+\mathcal{E}_{\nu}-\omega\right)\mathcal{X}_{\pi \nu}(\omega) + \delta\mathcal{H}_{\pi \nu}^{20}(\omega)=-\mathcal{F}_{\pi \nu}^{20}, \\
    \left(\mathcal{E}_{\pi}+\mathcal{E}_{\nu}+\omega\right)\mathcal{Y}_{\pi \nu}(\omega) + \delta\mathcal{H}_{\pi \nu}^{02}(\omega)=-\mathcal{F}_{\pi \nu}^{02}.
  \end{aligned}
\end{equation}
To solve the pnQFAM equations, one starts from initial induced densities $\mathcal{X}_{\pi\nu}(\omega)$ and $\mathcal{Y}_{\pi\nu}(\omega)$.
With known $\mathcal{X}_{\pi\nu}(\omega)$ and $\mathcal{Y}_{\pi\nu}(\omega)$, $ \delta\mathcal{H}_{\pi \nu}^{20}$
and $ \delta\mathcal{H}_{\pi \nu}^{02}$ can be computed through the connection with single-particle basis,
as shown in the following. Firstly, one obtains the induced densities in single-particle basis from the Bogoliubov transformation,
\begin{equation}
  \delta R_{pn}(\omega)
  = \left(\begin{matrix}
      \delta\rho^{(+)}_{p n}(\omega)        & \delta\kappa^{(+)}_{p n}(\omega)    \\
      -\delta\kappa^{(-)\ast}_{p n}(\omega) & -\delta\rho^{(-)\ast}_{p n}(\omega)
    \end{matrix}\right),
\end{equation}
where the induced density matrix $\delta\rho^{(\pm)}_{pn}(\omega)$
and induced pairing tensor $\delta\kappa^{(\pm)}_{pn}(\omega)$ read
\begin{equation}
  \begin{aligned}
    \delta \rho  ^{(+)}_{pn}(\omega) & = +\sum_{\pi\nu} \left[U     _{p \pi} \mathcal{X}     _{\pi\nu}(\omega) V     _{n \nu} - V^\ast_{p \pi} \mathcal{Y}     _{\pi\nu}(\omega) U^\ast_{n \nu}\right], \\
    \delta \rho  ^{(-)}_{pn}(\omega) & = -\sum_{\pi\nu} \left[V^\ast_{p \pi} \mathcal{X}^\ast_{\pi\nu}(\omega) U^\ast_{n \nu} - U     _{p \pi} \mathcal{Y}^\ast_{\pi\nu}(\omega) V     _{n \nu}\right], \\
    \delta \kappa^{(+)}_{pn}(\omega) & = +\sum_{\pi\nu} \left[U     _{p \pi} \mathcal{X}     _{\pi\nu}(\omega) U     _{n \nu} - V^\ast_{p \pi} \mathcal{Y}     _{\pi\nu}(\omega) V^\ast_{n \nu}\right], \\
    \delta \kappa^{(-)}_{pn}(\omega) & = -\sum_{\pi\nu} \left[V^\ast_{p \pi} \mathcal{X}^\ast_{\pi\nu}(\omega) V^\ast_{n \nu} - U     _{p \pi} \mathcal{Y}^\ast_{\pi\nu}(\omega) U     _{n \nu}\right].
  \end{aligned}
\end{equation}

As a consequence,
the induced single-particle Hamiltonian $\delta h^{(\pm)}_{pn}(\omega)$
and the induced pairing field $\delta \Delta^{(\pm)}_{pn}(\omega)$ can be calculated,
which form the induced Hamiltonian in the single-particle basis,
\begin{equation}
  \delta H_{pn} (\omega)
  = \left(\begin{matrix}
      \delta h     ^{(+)   }_{p n}(\omega)  & \delta\Delta^{(+) }_{p n}(\omega)    \\
      -\delta\Delta^{(-)\ast}_{p n}(\omega) & -\delta h   ^{(-)\ast}_{p n}(\omega)
    \end{matrix}\right).
\end{equation}
The details of $\delta h^{(\pm)}_{pn}(\omega)$ and $\delta \Delta^{(\pm)}_{pn}(\omega)$ are presented in the following subsections.
Finally, via the Bogoliubov transformation,
the elements of the induced Hamiltonian $\delta \mathcal{H}(\omega)$ read,
\begin{equation}
  \begin{aligned}
    \delta \mathcal{H}^{20}_{\pi\nu}(\omega)
    = \sum_{pn} & \left[ + U^\ast_{p \pi} \delta h     ^{(+)    }_{pn}(\omega) V^\ast_{n \nu}
    +                      U^\ast_{p \pi} \delta \Delta^{(+)    }_{pn}(\omega) U^\ast_{n \nu}\right.    \\
                & \left.           - V^\ast_{p \pi} \delta \Delta^{(-)\ast}_{pn}(\omega) V^\ast_{n \nu}
    -                      V^\ast_{p \pi} \delta h     ^{(-)\ast}_{pn}(\omega) U^\ast_{n \nu}\right],   \\
    \delta \mathcal{H}^{02}_{\pi\nu}(\omega)
    = \sum_{pn} & \left[ - V     _{p \pi} \delta h     ^{(+)    }_{pn}(\omega) U     _{n \nu}
    -                      V     _{p \pi} \delta \Delta^{(+)    }_{pn}(\omega) V     _{n \nu}\right.    \\
                & \left. +           U     _{p \pi} \delta \Delta^{(-)\ast}_{pn}(\omega) U     _{n \nu}
      +                      U     _{p \pi} \delta h     ^{(-)\ast}_{pn}(\omega) V     _{n \nu}\right].
  \end{aligned}
\end{equation}
Through Eq.~\eqref{eq:pnqfam}, new $\mathcal{X}_{\pi\nu}(\omega)$ and $\mathcal{Y}_{\pi\nu}(\omega)$ are obtained.
One can repeat this procedure, until it converges.
By introducing an infinitesimal imaginary part $\gamma$
in Eq.~\eqref{eq:pnqfam},
one can obtain the strength function $S(\hat{F},\omega)$ by,
\begin{equation}
  S(\hat{F},\omega) = -\frac{1}{\pi}\mathrm{Im}[R(\hat{F},\omega_\gamma)],
\end{equation}
where the response function $R(\hat{F},\omega_\gamma)$ with $\omega_\gamma = \omega +\mathrm{i}\gamma$ is given by
\begin{equation}
  R(\hat{F},\omega_\gamma)
  =\sum_{\pi\nu} \mathcal{F}^{20\ast}_{\pi\nu} \mathcal{X}_{\pi\nu}(\omega_\gamma)
  +              \mathcal{F}^{02\ast}_{\pi\nu} \mathcal{Y}_{\pi\nu}(\omega_\gamma).
\end{equation}
In the present paper, we concentrate on the Fermi (F) transitions and GT transitions,
which do not change the orbital motion. The operators are defined as,
\begin{equation}
  \begin{aligned}
    \hat{F}^{(\text{F})}  & = \hat{\tau}_-,                                \\
    \hat{F}^{(\text{GT})} & = \sum_{K=0,\pm1} \hat{\Sigma}_K \hat{\tau}_-,
  \end{aligned}
\end{equation}
where $\hat{\Sigma}$ is the relativistic spin operator.
Due to the axial symmetry, $K=\pm1$ give the same contribution to the response function.
To reproduce the strength function in the limit without pairing correlations,
the frequency $\omega$ is shifted by the chemical potential difference $\lambda_n-\lambda_p$.

\subsection{Induced single-particle Hamiltonian}
The ph interaction is described by
the relativistic density-dependent point-coupling density functional DD-PC1 \cite{Niksic_2008_DDPC1}.
In order to describe the GT transitions,
here we include the isovector-pseudovector interaction as \cite{Vale_2021_TPV}.
Thus, the total Lagrangian reads,
\begin{equation}
  \begin{aligned}
    \mathcal{L}= & \bar{\psi}(\mathrm{i} \gamma_\xi  \partial^\xi -m) \psi                                                                           \\
                 & -\frac{1}{2} \alpha_\text{S}(\bar{\psi} \psi)(\bar{\psi} \psi)
    -  \frac{1}{2} \alpha_\text{V}\left(\bar{\psi} \gamma^\xi \psi\right)\left(\bar{\psi} \gamma_\xi \psi\right)                                     \\
                 & -\frac{1}{2} \alpha_\text{TV}\left(\bar{\psi} \gamma^\xi \vec{\tau} \psi\right)\left(\bar{\psi} \gamma_\xi \vec{\tau} \psi\right) \\
                 & -\frac{1}{2} \delta_\text{S}\left(\partial_\xi \bar{\psi} \psi\right)\left(\partial^\xi \bar{\psi} \psi\right)
    -e \bar{\psi} \gamma^\xi  A_\xi \frac{\left(1-\tau_0\right)}{2} \psi                                                                             \\
                 & -\frac{1}{2} \alpha_\text{TPV}
    \left(\bar{\psi} \gamma^5 \gamma^\xi \vec{\tau} \psi\right)
    \left(\bar{\psi} \gamma_5 \gamma_\xi \vec{\tau} \psi\right),
  \end{aligned}
\end{equation}
where $\alpha_\text{S},\alpha_\text{V},\alpha_\text{TV}$, and $\alpha_\text{TPV}$
are the coupling constant of the isoscalar-scalar (S), isoscalar-vector (V),
isovector-vector (TV), and isovector-pseudovector (TPV) interaction, respectively.
$\delta_\text{S}$ is the coupling constant of the derivative term
which shows the density dependence.
$\vec{\tau}$ is the isospin Pauli matrix, the arrow indicates vectors in isospin space,
$\psi$ is the Dirac spinor field,
$\gamma^5$ and $\gamma^\xi$ are the Dirac gamma matrices, and $\xi$ represents a Minkowski index.

For the charge-exchange transitions,
only the TV and TPV interactions that are associated with isovector 4-currents
give a nonzero contribution to the induced single-particle Hamiltonian matrix $\delta h^{(\pm)}_{pn}$,
\begin{equation}\label{eq:dh_pn}
  \delta h^{(\pm)}_{pn}
  = \left\langle p \middle|
  \delta h^{(\pm)}_\text{TV}(\bm{r})+\delta h^{(\pm)}_\text{TPV}(\bm{r})
  \middle| n \right\rangle.
\end{equation}
Here, $\delta h^{(\pm)}_\text{TV}(\bm{r})$ and $\delta h^{(\pm)}_\text{TPV}(\bm{r})$ read,
\begin{equation}
  \begin{aligned}
    \delta h^{(\pm)}_\text{TV}(\bm{r})
     & = \alpha_\text{TV} \gamma_\xi \vec{\tau}
    \cdot \delta \vec{j}_\text{TV}^{(\pm)\xi}(\bm{r}),    \\
    \delta h^{(\pm)}_\text{TPV}(\bm{r})
     & = \alpha_\text{TPV} \gamma_5 \gamma_\xi \vec{\tau}
    \cdot \delta \vec{j}_\text{TPV}^{(\pm)\xi}(\bm{r}),
  \end{aligned}
\end{equation}
where the induced densities $\delta\vec{j}^\xi=(\delta\vec{\rho},\delta\vec{\bm{j}})$ are given by
\begin{equation}
  \begin{aligned}
    \delta \vec{\rho}^{(\pm)}_\text{TV}(\bm{r})
    = & \sum_{       l        l ^\prime}  \delta\rho^{(\pm)}_{       l         l ^\prime} \Phi^{\dagger}_{       l ^\prime}             \vec{\tau} \Phi_{       l }
    +                \sum_{\tilde{l}\tilde{l}^\prime}  \delta\rho^{(\pm)}_{\tilde{l} \tilde{l}^\prime} \Phi^{\dagger}_{\tilde{l}^\prime}             \vec{\tau} \Phi_{\tilde{l}},  \\
    \delta \vec{\bm{j}}^{(\pm)}_\text{TV}(\bm{r})
    = & \sum_{\tilde{l}        l ^\prime} \delta\rho^{(\pm)}_{\tilde{l}        l ^\prime} \Phi^{\dagger}_{       l ^\prime} \bm{\sigma} \vec{\tau} \Phi_{\tilde{l}}
    -                \sum_{       l  \tilde{l}^\prime} \delta\rho^{(\pm)}_{       l  \tilde{l}^\prime} \Phi^{\dagger}_{\tilde{l}^\prime} \bm{\sigma} \vec{\tau} \Phi_{       l } , \\
    \delta \vec{\rho}^{(\pm)}_\text{TPV}(\bm{r})
    = & - \mathrm{i} \sum_{\tilde{l}        l ^\prime} \delta\rho^{(\pm)}_{\tilde{l}        l ^\prime} \Phi^{\dagger}_{       l ^\prime}             \vec{\tau} \Phi_{\tilde{l}}
    +     \mathrm{i} \sum_{       l  \tilde{l}^\prime} \delta\rho^{(\pm)}_{       l  \tilde{l}^\prime} \Phi^{\dagger}_{\tilde{l}^\prime}             \vec{\tau} \Phi_{       l },  \\
    \delta \vec{\bm{j}}^{(\pm)}_\text{TPV}(\bm{r})
    = & -            \sum_{       l         l ^\prime} \delta\rho^{(\pm)}_{       l         l ^\prime} \Phi^{\dagger}_{       l ^\prime} \bm{\sigma} \vec{\tau} \Phi_{       l }
    -                \sum_{\tilde{l} \tilde{l}^\prime} \delta\rho^{(\pm)}_{\tilde{l} \tilde{l}^\prime} \Phi^{\dagger}_{\tilde{l}^\prime} \bm{\sigma} \vec{\tau} \Phi_{\tilde{l}}.
  \end{aligned}
\end{equation}
Here, $l$ and $\tilde{l}$ denote the indices of the axially deformed HO wavefunction
used to expand the large and small components, respectively,
which can be found in Ref.~\cite{Niksic_2014_DIRHB}.
Within the pnQFAM framework, only $\delta\rho^{(\pm)}_{pn}$ contribute to Eq.~\eqref{eq:dh_pn}.
The indices $p$ ($n$) run over all proton (neutron) single-particle states with projected total angular momentum $\Omega > 0$,
\begin{equation}
  \Phi_l(\bm{r})
  = \frac{1}{\sqrt{2 \pi}}
  \varphi_{n_z}(z,b_z)\varphi_{n_r}^{|\Lambda|} (r,b_r)
  \mathrm{e}^{\mathrm{i} \Lambda \phi}
  \chi_{1/2 m_s}\chi_{1/2 m_t},
\end{equation}
and their time-reversed states. The spatial part consists of the wave function along the symmetric axis $\varphi_{n_z}(z,b_z)$, and
the wave function perpendicular to the symmetric axis including $\varphi_{n_r}^{|\Lambda|} (r,b_r)$ and
the phase factor $\mathrm{e}^{\mathrm{i} \Lambda \phi}$.
$b_{z} (b_r)$ is the oscillator length in the $z$-($r$-) direction,
$n_r(n_z)$ is the number of nodes in the $z$-($r$-) direction,
$\Lambda$ is the projected orbital angular momentum,
$m_s(m_t)$ is the projected spin (isospin).
Consequently, the productions in densities and currents read,
\begin{equation}
  \begin{aligned}
      & \Phi^\dagger_l \Phi_{l^\prime}                                         \\
    = & \varphi_{n_z}(z,b_z)\varphi_{n_r}^{|\Lambda|} (r,b_r)
    \varphi_{n_z^\prime}(z,b_z)\varphi_{n_r^\prime}^{|\Lambda^\prime|} (r,b_r) \\
      & \frac{\mathrm{e}^{\mathrm{i} (-\Lambda + \Lambda^\prime) \phi}}{2 \pi}
    \chi_{1/2 m_s}^\dagger\chi_{1/2 m_s^\prime}
    \chi_{1/2 m_t}^\dagger\chi_{1/2 m_t^\prime},                               \\
      & \Phi^\dagger_l \hat{\bm{\sigma}}\hat{\bm{\tau}}\Phi_{l^\prime}         \\
    = & \varphi_{n_z}(z,b_z)\varphi_{n_r}^{|\Lambda|} (r,b_r)
    \varphi_{n_z^\prime}(z,b_z)\varphi_{n_r^\prime}^{|\Lambda^\prime|} (r,b_r) \\
      & \frac{\mathrm{e}^{\mathrm{i} (-\Lambda + \Lambda^\prime) \phi}}{2 \pi}
    \chi_{1/2 m_s}^\dagger\hat{\bm{\sigma}}\chi_{1/2 m_s^\prime}
    \chi_{1/2 m_t}^\dagger\hat{\vec{\tau}}\chi_{1/2 m_t^\prime}.
  \end{aligned}
\end{equation}
The angular part $\mathrm{e}^{\mathrm{i} (-\Lambda + \Lambda^\prime) \phi}$
can be canceled in the calculation of induced single-particle Hamiltonian matrix elements.
The spin and isospin parts are determined by the Clebsch-Gordan coefficients,
\begin{equation}
  \begin{aligned}
    \left\langle \tfrac{1}{2} m_s \middle| \hat{\bm{\sigma}}_{\Delta m_s} \middle| \tfrac{1}{2} m_s^\prime\right\rangle
    = & \sqrt{3}
    \left(\tfrac{1}{2} m_s 1 \Delta m_s \middle| \tfrac{1}{2} m_s^\prime\right), \\
    \left\langle \tfrac{1}{2} m_t \middle| \hat{\vec{\tau}}_{\Delta m_t} \middle| \tfrac{1}{2} m_t^\prime\right\rangle
    = & \sqrt{3}
    \left(\tfrac{1}{2} m_t 1 \Delta m_t \middle| \tfrac{1}{2} m_t^\prime\right).
  \end{aligned}
\end{equation}

\subsection{Induced pairing field}
The finite-range separable pairing force \cite{YTian_2009_axialpairing,YTian_2009_pairing,YTian_2009_PRC}
is adopted for pp interaction in pnQFAM.
By introducing the separable pairing interaction,
one can avoid to calculate the two-body pp interaction matrix elements hence reducing computational costs.
The induced pairing field $\delta\Delta^{(\pm)}_{pn}$ can be obtained from
the induced pairing tensor $\delta\kappa^{(\pm)}_{pn}$,
\begin{equation}
  \delta\Delta^{(\pm)}_{pn} = \sum_{p^\prime n^\prime} \bar{v}^\text{pp}_{p n p^\prime n^\prime} \delta\kappa^{(\pm)}_{p^\prime n^\prime},
\end{equation}
with the pp interaction matrix element,
\begin{equation}
  \bar{v}^\text{pp}_{p n p^\prime n^\prime}
  = \left\langle p n \middle| V^\mathrm{pp} \middle| p^\prime n^\prime \right\rangle
  - \left\langle p n \middle| V^\mathrm{pp} \middle| n^\prime p^\prime \right\rangle.
\end{equation}

The separable pp interaction \cite{YTian_2009_pairing,YTian_2009_axialpairing,YTian_2009_PRC}
reads,
\begin{equation}
  \begin{aligned}
      & V^\mathrm{pp} (\bm{r}_1, \bm{r}_2, \bm{r}_1^\prime, \bm{r}_2^\prime )                          \\
    = & - G \delta(\bm{R}-\bm{R}^\prime) P(r,z) P(r^\prime,z^\prime) \frac{1}{2}(1 - P^\sigma P^\tau),
  \end{aligned}
\end{equation}
where
$\bm{R} = \left(\bm{r}_1 + \bm{r}_2\right)/2$ and $\bm{r} = \bm{r}_2 - \bm{r}_1$
correspond to the center-of-mass and relative coordinates, respectively.
In the cylindrical coordinates,
$P(r,z)=\mathrm{e}^{-\frac{r^2+z^2}{4a^2}}/(4\pi a^2)^\frac{3}{2}$. The strength $G$ and the range $a$ determine the pairing force.
$P^\sigma$ and $P^\tau$ are the exchange operators for the spin and isospin, respectively.
It's more convenient to calculate the matrix elements in spin and isospin coupled representation, by
\begin{equation}
  \begin{aligned}
      & \left| \tfrac{1}{2}~m_t~\tfrac{1}{2}~-m_t \right\rangle                                                       \\
    = & \sum_{T=0,1} \left( \tfrac{1}{2}~m_t~\tfrac{1}{2}~-m_t \middle| T~m_T=0 \right) \left| T~m_T=0 \right\rangle, \\
      & \left| \tfrac{1}{2} m_{s_1} \tfrac{1}{2} m_{s_2} \right\rangle                                                \\
    = & \sum_{\substack{S=0,1,                                                                                        \\m_S =-S,S}} \left( \tfrac{1}{2}~m_{s_1}~\tfrac{1}{2}~m_{s_2} \middle| S~m_S \right) \left| S, m_S   \right\rangle,
  \end{aligned}
\end{equation}
where $S$ and $T$ are the total spin and isospin of the pair,
$m_S$ and $m_T$ are the corresponding projections.
By applying $P^\sigma P^\tau$ on the coupled state,
\begin{equation}
  P^\sigma P^\tau \left| S m_S, T m_T \right\rangle
  = (-1)^{S+T} \left| S m_S, T m_T \right\rangle,
\end{equation}
one can find that only $S+T=1$ pairs are allowed in the pp interaction,
which can be sorted into isovector pairs ($S=0,T=1$) and isoscalar pairs ($S=1,T=0$).

In the RHB model,
neutron-neutron pairs and proton-proton pairs are considered.
As a consequence, only isovector pairs exist in the RHB model.
However, in the charge-exchange QFAM framework which involves $\delta\kappa^{(\pm)}_{pn}$,
both isovector and isoscalar pairs are allowed.

The pp matrix elements are calculated
in the axially deformed HO basis,
\begin{equation}
  V^\mathrm{pp}_{l_1 l_2,l_1^\prime l_2^\prime}
  =\left\langle l_1 l_2 \right|
  V^\mathrm{pp} (\bm{r}_1, \bm{r}_2, \bm{r}_1^\prime, \bm{r}_2^\prime )
  \left|l_1^\prime l_2^\prime\right\rangle.
\end{equation}

The spatial part of the pairing matrix element can be calculated in the center-of-mass frame. With the use of
one- and two-dimensional Talmi-Moshinsky brackets
\cite{YTian_2009_axialpairing,Niksic_2014_DIRHB,Bjelcic_2020_QFAM,Ravlic_2024_DpnRQRPA}, it
gives,
\begin{gather}
  \begin{aligned}
         & \varphi_{n_{z 1}}(z_1,b_z) \varphi_{n_{z 2}}(z_2,b_z)                                \\
    =    & \sum_{N_z=0}^{+\infty} \sum_{n_z=0}^{+\infty}
    M_{N_z, n_z}^{n_{z_1}, n_{z 2}} \varphi_{N_z}(Z,b_z/\sqrt{2}) \varphi_{n_z}(z,\sqrt{2}b_z), \\
         & \varphi_{n_{r_1}}^{|\Lambda_1|}(r_1,b_r)
    \varphi_{n_{r 2}}^{|\Lambda_2|}(r_2,b_r)                                                    \\
    = {} & \sum_{M,m=-\infty}^{+\infty} \sum_{N_r,n_r=0}^{+\infty}                              \\
         & M_{N_r \Lambda n_r \lambda}^{n_{r_1} \Lambda_1 n_{r_2} \Lambda_2}
    \varphi_{N_r}^\Lambda(R,b_r/\sqrt{2})
    \varphi_{n_r}^\lambda(r,\sqrt{2}b_r) ,
  \end{aligned}
\end{gather}
which requires the following selection rules,
\begin{equation}
  \begin{aligned}
    n_{z1} + n_{z2}                                 & = N_z + n_z,                                 \\
    2 n_{r1} + |\Lambda_1| + 2 n_{r2} + |\Lambda_2| & = 2 N_{r} + |\Lambda|  + 2 n_r + |\lambda| , \\
    \Lambda_1 + \Lambda_2                           & = \Lambda + \lambda.
  \end{aligned}
\end{equation}

The spin-isospin part of pairing matrix elements can be given by the Clebsch-Gordan coefficients
according to the total spin $S$ and total isospin $T$.

Analogously to Ref.~\cite{YTian_2009_axialpairing},
it's convenient to write the matrix elements in terms of separable terms,
\begin{equation}
  \bar{v}^\text{pp}_{1 2,1^\prime 2^\prime}
  = - G \sum_{T=0,1} f(T) \sum_{N_z N_r} W^{N_z,N_r,T}_{1 2} W^{N_z,N_r,T}_{1^\prime 2^\prime}.
\end{equation}
Here, the factor $f(1)=1$, and $f(0)=f_\text{IS}$ is introduced to adjust the isoscalar pairing strength,
as the isoscalar strength can not be constrained in the ground state \cite{Vale_2021_TPV}.
The separable term can be further decomposed into two parts,
\begin{equation}
  W^{N_z,N_r,T=0(T=1)}_{1 2} = W^{N_z,N_r}_{1 2} \times W^{T=0(T=1)}_{1 2},
\end{equation}
where the spatial part reads,
\begin{equation}
  \begin{aligned}
      & W^{N_z,N_r}_{1 2}                                                                                                                       \\
    = & \frac{b_r \sqrt{b_z}}{(2 \pi)^{3 / 4}} M_{N_z n_z}^{n_{z_1} n_{z_2}} \frac{\sqrt{n_z !}}{2^{\frac{n_z}{2}}\left(\frac{n_z}{2}\right) !}
    \frac{\left(a^{2}-b_z^{2}\right)^{\frac{n_z}{2}}}{\left(a^{2}+b_z^{2}\right)^{\frac{n_z+1}{2}}}                                             \\
      & \times M_{N_r \Lambda n_r 0}^{n_{r_1} \Lambda_1 n_{r_2} \Lambda_2}
    \frac{\left(b_r^{2}-a^{2}\right)^{n_{r}}}{\left(b_r^{2}+a^{2}\right)^{n_{r}+1}},
  \end{aligned}
\end{equation}
and the spin-isospin part reads,
\begin{equation}
  \begin{aligned}
    W^{T=1}_{1 2} = & \frac{\delta_{m_{s_1},-m_{s_2}}(-1)^{m_{s_1}-\tfrac{1}{2}}}{\sqrt{2}}, \\
    W^{T=0}_{1 2} = & \frac{\delta_{m_{s_1},-m_{s_2}}}{\sqrt{2}} + \delta_{m_{s_1},m_{s_2}}.
  \end{aligned}
\end{equation}
With the separable terms, the induced pairing field $\delta\Delta^{(\pm)}_{pn}$ can be written as,
\begin{equation}
  \delta\Delta^{(\pm)}_{pn} = -G \sum_{T=0,1} F(T) \sum_{N_z N_r} W^{N_z,N_r,T}_{p n} P^{N_z,N_r,T,(\pm)},
\end{equation}
with the definition,
\begin{equation}
  P^{N_z,N_r,T,(\pm)} = \sum_{p n} W^{N_z,N_r,T}_{p n} \delta\kappa^{(\pm)}_{p n}.
\end{equation}

\section{Numerical Details\label{sec:num}}

In this work, all calculations are performed using the DD-PC1 parameter set \cite{Niksic_2008_DDPC1}
combining with the corresponding isovector-pseudovector strength $\alpha_{\text{TPV}} = -0.734\text{ fm}^2$
adopted from Ref.~\cite{Vale_2021_TPV} for the ph channel.
The pp channel is treated using a finite-range separable pairing interaction \cite{YTian_2009_axialpairing}
including both isovector and isoscalar channels.
Unless otherwise specified, in the following, the isoscalar pairing strength
is set to the same value with the isovector one, e.g. $f_\text{IS}=1$.

The imaginary part $\gamma$ of the excitation frequency
corresponds to a Lorentzian smearing with a full width at half maximum (FWHM) of $2\gamma$.
As is well known, giant resonances exhibit considerable width
due to the damping mechanisms.
Parts of these damping effects, however,
cannot be captured within the QRPA or QFAM frameworks.
Improvements to the treatment of damping can be achieved
by incorporating higher-order many-body correlations,
such as two-particle-two-hole (2p-2h) configurations
\cite{Gambacurta_2020_PRL, CLBai_2010_PRL}
or (quasi)particle-vibration coupling [(Q)PVC] effects
\cite{YFNiu_2015_PRL, YFNiu_2018_PLB, ZZLi_2023_PRL, ZZLi_2024_PRC, Litvinova_2019_PRC}.
However, such extensions involve extremely high computational cost
and are mostly limited to spherical systems.
Taking advantage of the computational efficiency of the QFAM framework,
QPVC effects have recently been included in axially deformed nuclei
for non-charge-exchange excitations based on the relativistic DFT \cite{YNZhang_2022_PRC},
and for charge-exchange excitations using the non-relativistic DFT \cite{QQLiu_2024_PRC}.
The present study serves as a foundation for future developments
aimed at incorporating both QPVC effects and deformation effects into charge-exchange transitions
within the relativistic DFT framework.
In the present work, QPVC effects are not included.
Instead, a relatively small fixed smearing width of $\gamma = 0.25\text{ MeV}$
is employed to enhance the resolution of individual peaks in the strength distributions.

\begin{figure}
  \center
  \includegraphics[scale=0.8]{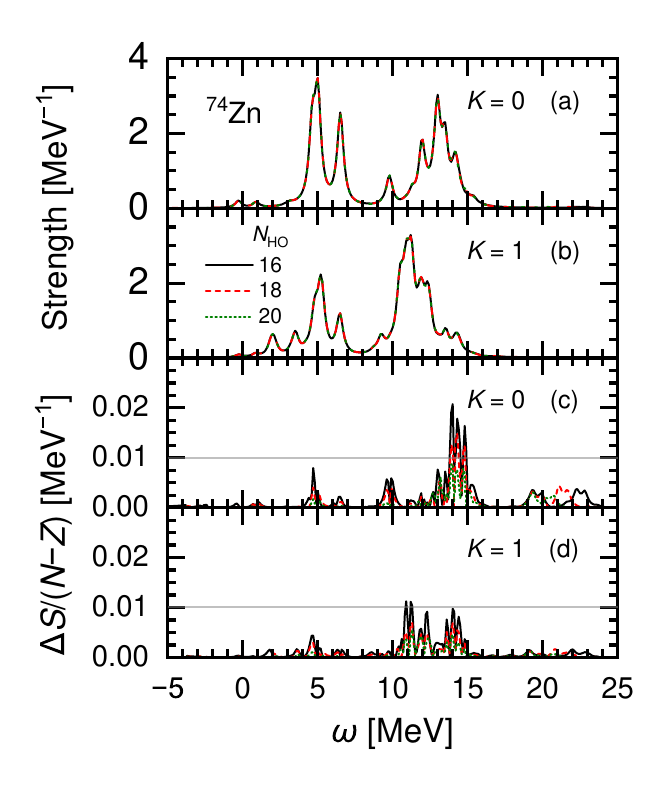}
    \caption{Gamow-Teller (GT) strength distributions in $^{74}\text{Zn}$ calculated 
    using different harmonic oscillator (HO) basis truncations ($N_\text{HO}=16$, 18, 20), 
    and the corresponding difference compared to the reference case ($N_{\text{HO}}^\text{max}=22$). 
    Panels (a) and (b) show the GT strength distributions for $K=0$ and $K=1$ modes, respectively,
    while panels (c) and (d) present the corresponding normalized differences relative to the reference case, scaled by the Ikeda sum rule ($N-Z$) for each $K$ component. 
    The truncations $N_\text{HO}=16$ (black solid), 18 (red dashed), and 20 (green dotted) are shown with distinct line styles.
    The gray horizontal line marks the 1\% level of $\delta S/(N-Z)$.\label{fig1}}
\end{figure}

The RHB and relativistic pnQFAM equations are numerically solved
by expanding in a finite axially deformed HO basis,
with the basis size truncated by the principal quantum number $N_\text{HO}$.
The value of $N_\text{HO}$ is chosen to ensure the convergence of the calculated strength functions.

Fig.~\ref{fig1} shows strength distributions in $^{74}\text{Zn}$ calculated using different harmonic oscillator (HO) basis truncations ($N_\text{HO}=16$, 18, 20), 
and the corresponding difference compared to the reference case ($N_{\text{HO}}^\text{max}=22$). 
The ground-state deformation converges to $\beta_2 \approx 0.186$ in all cases,
thereby minimizing any impact from deformation changes due to the basis sizes.

The strength difference is defined as $\delta S = |S(N_\text{HO}) - S(N_{\text{HO}}^\text{max})|$
and normalized by $N - Z$ to facilitate comparison with the Ikeda sum rule for each $K$ component.
Panels (a) and (b) of Fig.~\ref{fig1} present the strengths for the $K = 0$ and $K=1$ component, respectively.
Panels (c) and (d) show the corresponding normalized differences relative to the reference case, scaled by the Ikeda sum rule ($N-Z$) for each $K$ component.
The strength differences obtained with $N_\text{HO} = 16$, $18$, and $20$ are indicated by
solid black, dashed red, and dotted green lines, respectively.
The gray horizontal line marks $1\%$ of the normalized value.
As shown in the figure,
within the energy range $\omega < 17\text{ MeV}$, where the majority of the strength is concentrated,
the strength differences exhibit stable peak positions and decrease with increasing $N_\text{HO}$.
When $N_\text{HO}=20$, the maximum differences fall below $1\%$ for both $K = 0$ and $K = 1$ components.
At higher excitation energies, the strength differences are  much smaller than $1\%$,
and slight energy shifts in peak positions are observed.
The above observations indicate that $N_\text{HO}=20$ gives well converged results, and therefore,
a basis size of $N_\text{HO}=20$ is adopted in all subsequent calculations.

\begin{figure}
  \center
  \includegraphics[scale=0.8]{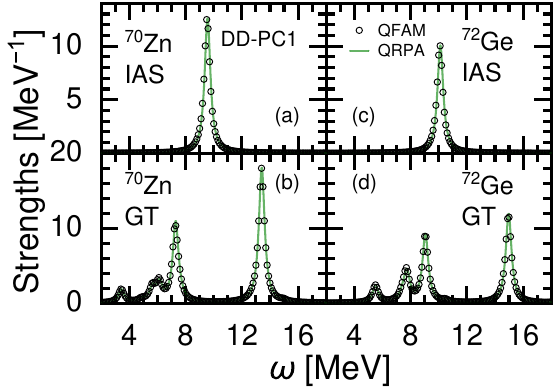}
  \caption{Comparisons of strength distributions between QFAM
    with $\beta_2=0.0$ (open black circles) and spherical QRPA  (solid green lines) calculations.
    The upper panels display the isobaric analog state (IAS) strength distributions,
    while the lower panels show the GT strength distributions.
    The left and right panels correspond to nuclei $^{70}\text{Zn}$ and $^{72}\text{Ge}$, respectively.
    \label{fig2}}
\end{figure}

To validate pnQFAM results,
results obtained using the QFAM approach are compared with
those from the QRPA approach as shown in Fig.~\ref{fig2}.
These calculations are performed using the same numerical parameters,
except for the assumed spatial symmetry:
the QRPA calculations are performed under the spherical symmetric assumption,
while the QFAM calculations assume the axial symmetry.
To ensure a consistent comparison, predicted spherical nuclei $^{70}\text{Zn}$ and $^{72}$Ge are selected for the comparison.
The corresponding isobaric analog state (IAS) strengths are shown in the upper panels (a) and (c),
and GT strength distributions are displayed in panels (b) and (d).
Open black circles represent QFAM results, while solid green lines correspond to QRPA results.
The two approaches exhibit excellent agreement for both IAS and GT strength distributions,
confirming the correctness of the QFAM implementation.

\section{Results and Discussion\label{sec:res}}
In this section, the relativistic pnQFAM approach is applied to study charge-exchange transitions in Zn isotopes, with a particular focus on
the deformation effects.

\begin{figure}
  \center
  \includegraphics[scale=0.8]{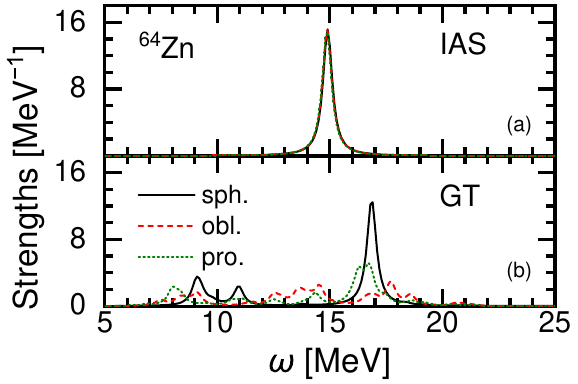}
  \caption{Comparisons of IAS and GT strengths for  different shapes of $^{64}\text{Zn}$ .
    The upper panel (a) displays the IAS strengths,
    while the lower panel (b) shows the GT strengths.
    Results based on spherical, prolate, and oblate configurations are indicated by
    solid black, dashed red, and dotted green lines, respectively.
    \label{fig3}}
\end{figure}

\paragraph*{Deformation effects on IAS and GT strengths.}
Comparisons of IAS and GT strengths for different shapes of $^{64}\text{Zn}$ are shown in Fig.~\ref{fig3}.
The oblate configuration corresponds to the ground state at $\beta_2 = -0.246$ with a total energy of $E = -555.719 \text{ MeV}$,
while the prolate configuration corresponds to the local minimum at $\beta_2 = 0.239$ with $E = -555.455 \text{ MeV}$.
The spherical configuration is obtained through a constrained calculation at $\beta_2 = 0.0$.
The results for spherical, prolate, and oblate configurations are plotted
using solid black, dashed red, and dotted green lines, respectively.
Panel (a) displays IAS strengths for the different configurations.
It can be seen that nuclear deformation has a rather small impact on the IAS strength distribution in this isotope,
with only slight negative shifts in the peak energies.
This observation is consistent with previous studies \cite{Rocamaza_2020_PRC}.
The weak deformation dependence arises from the fact that
the IAS energy is approximately proportional to the factor $1 - \beta_2^2 / 4\pi$ \cite{Rocamaza_2020_PRC}.
Since the quadrupole deformation $\beta_2$ is typically small,
the correction term contributes only a few percent, resulting in minor energy shifts.

Panel (b) displays the GT strengths for the different configurations.
In contrast to the IAS case, the GT strength exhibits a strong dependence on nuclear deformation.
For low-lying excitations around $10\text{ MeV}$,
these peaks are shifted toward lower energies.
Meanwhile, the GTR peak is largely fragmented.
Moreover, the oblate configuration leads to more pronounced fragmentation compared to the prolate case.
This fragmentation arises from the single-particle level splitting in deformed nuclei,
which largely increases the configuration space of 2qp pairs.

\begin{figure*}
  \center
  \includegraphics[scale=0.75]{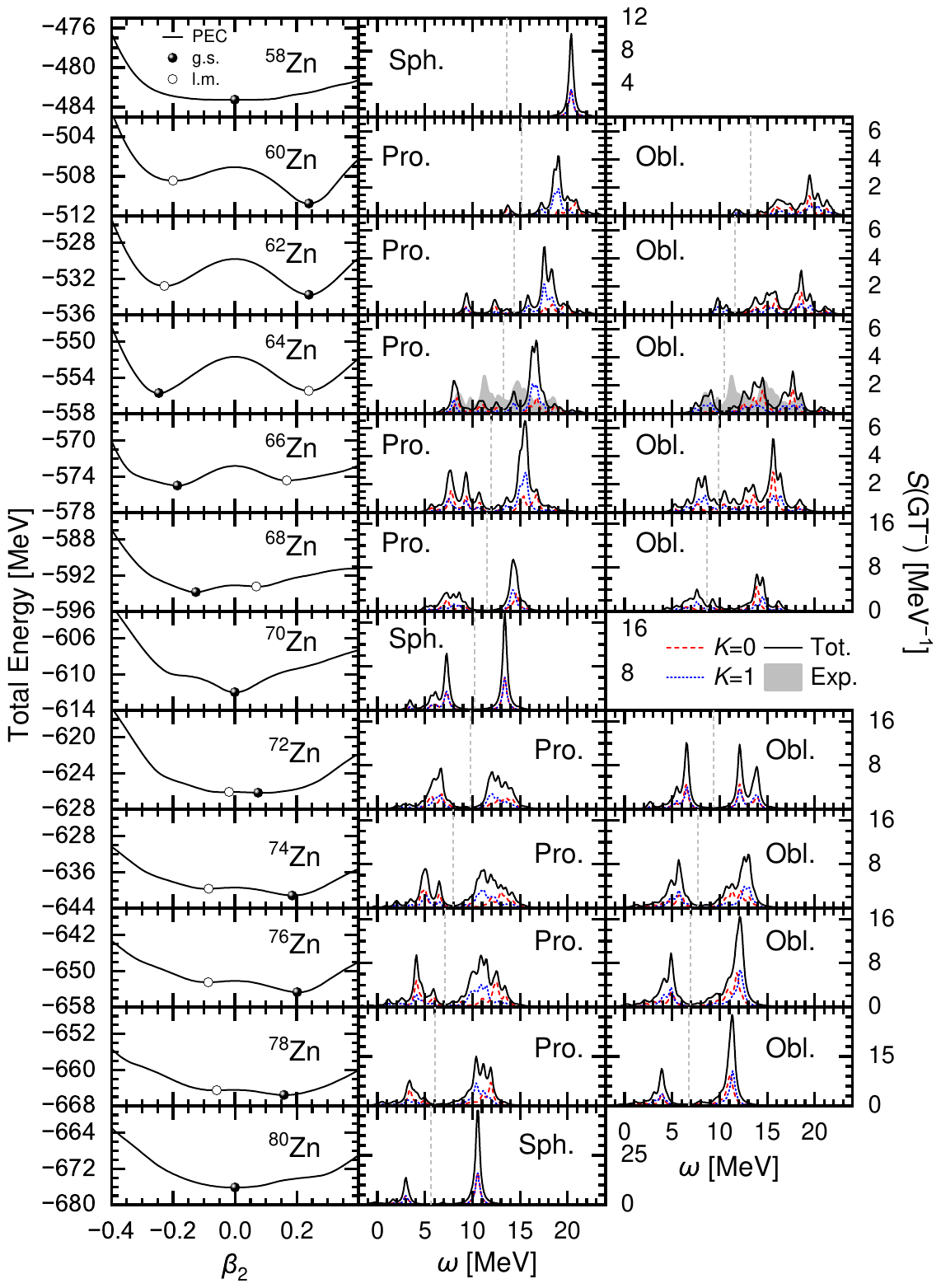}
  \caption{Left Column:
    Potential energy curves for Zn isotopes,
    showing the total energy (black solid lines) as a function of the quadrupole deformation parameter $\beta_2$.
    Global and local minima are indicated by black spheres and open circles, respectively.
    The middle and right Columns:
    GT strengths based on the configurations marked in the left column.
    The total GT strength is shown by black solid lines,
    with the $K=0$ and $K=1$ components represented by red dashed and blue dotted lines, respectively.
    For $^{64}\text{Zn}$,
    the experimental $B(\text{GT})$  is smeared with a width of $0.25\text{ MeV}$
    and scaled by a factor of $4.67$ (gray shaded area) according to the ratio of theoretical and experimental summed strengths \cite{FDiel_2019_PRC}.
    The vertical gray dashed line indicates the lower boundary of the GT resonance (GTR) region.
    \label{fig4}}
\end{figure*}

\paragraph*{Deformation effects on GT strengths in Zn isotopes.}
Considering that IAS strengths are almost unaffected by deformation,
in the following, the deformation effects on GT strengths are focused on.
A systematic investigation for Zn isotopes, ranging from $^{58}\text{Zn}$ to $^{80}\text{Zn}$, are performed.
The left column in Fig.~\ref{fig4} presents the potential energy curves (PECs) for Zn isotopes
with global and local minima indicated by black spheres and open circles, respectively.
These black curves show the evolution of total energies
as a function of the quadrupole deformation parameter $\beta_2$.

In $^{58}\text{Zn}$ with $N=28$, the neutron shell closure favors a spherical shape.
As neutrons are progressively added and begin to occupy orbitals originating from the $2p_{3/2}$ and $1f_{5/2}$ spherical shells,
the PECs of $^{60\text{-}68}\text{Zn}$ exhibit two minima, one on the prolate and the other on the oblate side.
For $^{60}\text{Zn}$, the prolate minimum at $\beta_2 = 0.240$ corresponds to the ground state with $E = -510.743$ MeV,
while a local oblate minimum emerges at $\beta_2 = -0.200$ with $E = -508.432$ MeV.
In $^{62}\text{Zn}$, the prolate ground state persists at $\beta_2 = 0.240$ ($E = -533.788$ MeV),
but the oblate local minimum becomes slightly deeper ($\beta_2 = -0.228$, $E = -532.791$ MeV),
indicating a reduced energy difference between the two configurations compared to $^{60}\text{Zn}$.
For $^{64}\text{Zn}$ which exhibits the largest deformation along the Zn chain,
the oblate minimum lies slightly lower in energy than the prolate minimum.
Beyond $^{64}\text{Zn}$, the deformation gradually decreases with the neutron number increasing towards the next sub-closed shell $N=40$.
The nucleus $^{70}\text{Zn}$ returns to a spherical shape as the neutrons fill the $2p_{1/2}$ shell that forms the sub-closed shell $N=40$.
For the heavier isotopes $^{72\text{-}78}\text{Zn}$,
the occupation of the $1g_{7/2}$ shell once again gives rise to two minima in the PECs.
In these cases, the ground state tends to favor a prolate shape, with a local oblate minimum also present.
Among them, $^{76}\text{Zn}$ with neutrons locating in the middle of $1g_{7/2}$ shell exhibits the largest deformation in this neutron-rich region.
The nucleus $^{80}\text{Zn}$ becomes spherical again with the closed shell $N=50$.

The corresponding GT strengths for the ground states and local minima
are displayed in the middle and right columns in Fig.~\ref{fig4}.
The middle column presents the strengths for spherical or prolate configurations,
while the right column shows those for oblate configurations.
The $K=0$, $K=1$ components, and the total GT strengths are plotted as red dashed, blue dotted lines, and black solid lines respectively.
For $^{64}\text{Zn}$, the experimental $B(\text{GT})$ data (gray shaded area) are smeared with a Lorentzian of width $0.25\text{ MeV}$
and scaled by a factor of $4.67$ according to the ratio of theoretical and experimental summed strengths \cite{FDiel_2019_PRC}.

The GT strength distributions include both low-lying states and the high-energy GTRs.
The high-energy GTRs correspond to spin-flip transitions ($j_> \rightarrow j_<$), 
while the low-energy states relate to
core-polarization ($j_> \rightarrow j_>$ or $j_< \rightarrow j_<$) and back spin-flip ($j_< \rightarrow j_>$) excitations \cite{Paar_2004_pnRQRPA}.
Following this spirit, the boundaries between low-lying and resonance regions are determined and shown by the vertical gray dashed lines.
It should mention that in our case $j$ is not a good quantum number due to the violation of spherical symmetry.
The corresponding $j_>$ or $j_<$ spherical shells only label the spherical origins of the single-particle levels. 

For total GT strengths, in spherical nuclei, $^{58}\text{Zn}$, $^{70}\text{Zn}$, and $^{80}\text{Zn}$,
the $K=0$ and $K=1$ components are degenerate, yielding concentrated strength distributions.
In contrast, for deformed nuclei ($^{60\text{-}68}\text{Zn}$ and $^{72\text{-}78}\text{Zn}$),
two components are split, yielding a broader and more fragmented distribution.
The degree of splitting increases with the deformation.
In $^{60\text{-}68}\text{Zn}$,
the oblate configurations generally exhibit more fragmented GT strength distributions
compared to the prolate configurations,
even when the absolute values of deformation parameter are similar.
However, in $^{72\text{-}78}\text{Zn}$,
the prolate configurations tend to show more pronounced fragmentation,
which can be attributed to the small deformations of oblate local minima with $\beta_2<-0.1$.

\begin{figure}
  \center
  \includegraphics[scale=0.8]{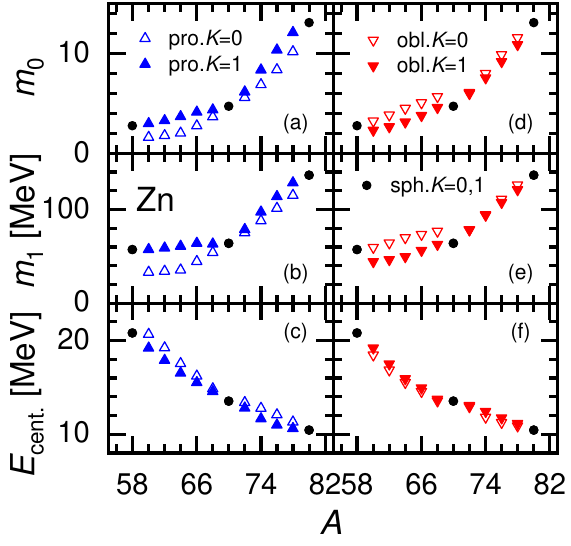}
  \caption{
    Non-energy-weighted summed strength  $m_0$ [panels (a) and (d)]
    energy-weighted summed strength  $m_1$  [panels (b) and (e)],
    and centroid energies $E_\text{cent.} = m_1/m_0$ [panels (c) and (f)]
    in the resonance region for the $K=0$ and $K=1$ strengths in Zn isotopes.
    For spherical nuclei, the $K=0$ and $K=1$ components are degenerated and represented by solid black circles.
    For deformed nuclei, open symbols denote the $K=0$ components, while solid symbols correspond to the $K=1$ components.
    Blue upward triangles and red downward triangles represent results obtained on top of prolate (left column) and oblate configurations (right column) , respectively.
    \label{fig5}}
\end{figure}

Additionally, for $K=0$ and $K=1$ strengths in resonance regions,
peaks of the $K=0$ components in prolate nuclei consistently locate at higher energies than the $K=1$ components,
whereas the reversed trend can be observed for oblate nuclei.
Moreover, in $^{60\text{-}68}\text{Zn}$,
the $K=0$ component in the prolate configurations exhibits a smaller strength than the $K=1$ component,
while the opposite is true for the oblate side.
This contrast disappears in $^{72\text{-}78}\text{Zn}$, where two components exhibit similar strengths.
These features between two components will be further investigated quantitatively via summed strengths and centroid energies in GTR region as presented in Fig.\ref{fig5}.

\paragraph*{Systematics of $K$-splitting in summed strengths and centroid energies in GTR region.}
In Fig.~\ref{fig5}, the non-energy-weighted summed strengths $m_0$ and energy-weighted summed strengths $m_1$,
as well as the corresponding centroid energies $E_\text{cent.} = m_1/m_0$,
for the $K=0$ and $K=1$ strengths in the resonance region in Zn isotopes are focused on.
For spherical nuclei, the $K=0$ and $K=1$ components are degenerated and represented by solid black circles.
For deformed nuclei, these quantities are displayed as open symbols for the $K=0$ components and solid symbols for the $K=1$ components.
Blue upward triangles represent prolate cases (left column), whereas red downward triangles indicate oblate cases (right column).

One can observe that both $m_0$ and $m_1$ exhibit overall increasing trends with the mass number.
A noticeable change in the slope occurs beyond $^{70}\text{Zn}$,
which is attributed to the occupation of neutron orbitals originating from the $1g_{9/2}$.
These additional neutrons enable new transitions contributing to the resonance region.
In spherical nuclei, $^{58}\text{Zn}$, $^{70}\text{Zn}$, and $^{80}\text{Zn}$,
the $K=0$ and $K=1$ components are the same due to the isotropy.
In contrast, for deformed nuclei, $^{60\text{-}68}\text{Zn}$ and $^{72\text{-}78}\text{Zn}$,
the summed strengths in GTR region exhibit deformation splitting:
in prolate configurations, the $K=1$ components of both $m_0$ and $m_1$ are consistently larger than the $K=0$ components,
while the opposite trend is observed in oblate configurations.

\begin{figure}
  \center
  \includegraphics[scale=0.8]{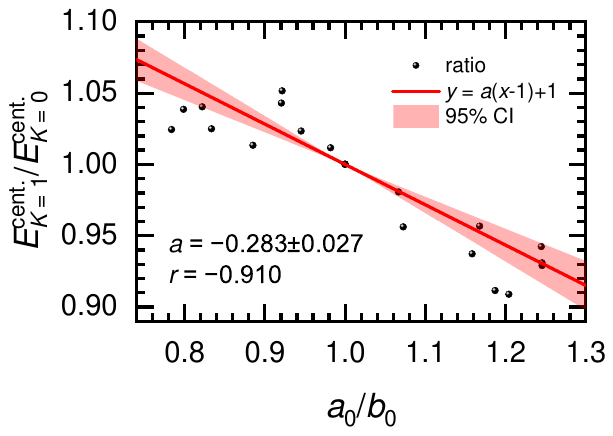}
  \caption{Relation between the centroid energy ratio of the $K=0$ and $K=1$ components in the GTR region
    and the axis ratio $a_0/b_0=\frac{1+\sqrt{\frac{5}{4\pi}}\beta_2}{1-\sqrt{\frac{5}{4\pi}}\frac{\beta_2}{2}}$.
    Ratios extracted from microscopic results are indicated by black spheres.
    The red line shows the fitted function,
    and the red shaded area represents the $95\%$ confidence interval (CI).
    \label{fig6}}
\end{figure}

Furthermore, the $m_0$ differences between two components in $^{60\text{-}68}\text{Zn}$
are comparable with those in $^{72\text{-}78}\text{Zn}$. However for the $m_1$ differences,
they are larger in $^{60\text{-}68}\text{Zn}$ compared to in $^{72\text{-}78}\text{Zn}$.
This reduction in $m_1$ splitting for heavier isotopes can be attributed to lower centroid energies in heavier nuclei.
Indeed, the centroid energies $E_\text{cent.}$ exhibit a smooth decreasing trend with increasing mass number.
Moreover, the $E_\text{cent.}$ splitting shows opposite trends in the prolate and oblate cases:
the $K=1$ ($K=0$) centroid energies are larger than the other ones in prolate (oblate) configurations, 
and the energy splitting shows an approximate proportionality to the nuclear deformation.
This behavior is similar with that observed in isovector giant dipole resonance (IVGDR) \cite{CChen_2025_arXiv},
and will be analyzed in Fig.~\ref{fig6}.

\begin{figure*}
  \center
  \includegraphics[width=0.9\textwidth]{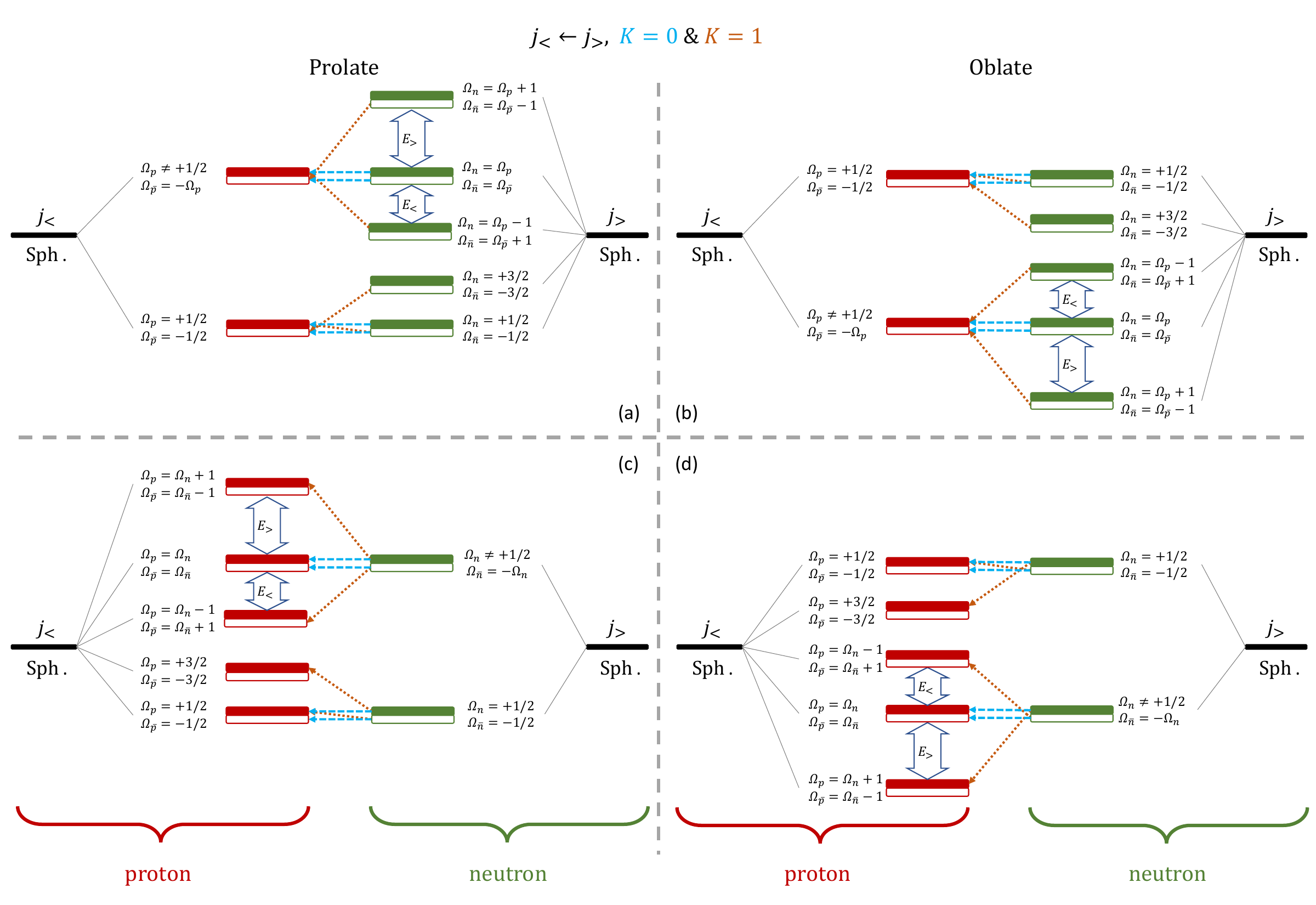}
  \caption{
    Schematic diagram illustrating $K=0$ and $K=1$ modes of GTR in the $\beta^-$ direction.
    The upper panels show transitions to specific final states,
    and the lower panels present transitions from specific final states.
    The left (right) panels illustrate the prolate (oblate) case.
    Proton orbitals (red bars) originate from $j_<$ spherical shell (black bars),
    while neutron orbitals (green bars) originate from $j_>$ spherical shell.
    Solid (open) bars denote states with positive projected angular momentum $\Omega_{p(n)}>0$
    (time-reversed states $\Omega_{\bar{p}(\bar{n})}<0$).
    Blue dashed and orange dotted arrows indicate transitions allowed by $K=0$ and $K=1$ selection rules, respectively.
    In the prolate case, the splitting energy between $\Omega+1$ and $\Omega$ levels
    is larger than that between $\Omega$ and $\Omega-1$ levels, and
    the opposite holds for the oblate case, which is reflected in the schematic drawings.  The single-particle transition energies allowed by  $K=0$ selection rules are set to be zero as benchmarks.
    \label{fig7}}
\end{figure*}

For the IVGDR, Ref.~\cite{Danos_1958_NP} established a relation between the centroid energy ratio of the two components
and the axis ratio $a_0/b_0=\frac{1+\sqrt{\frac{5}{4\pi}}\beta_2}{1-\sqrt{\frac{5}{4\pi}}\frac{\beta_2}{2}}$ related to the quadrupole deformation parameter $\beta_2$.
Following the same approach, Fig.~\ref{fig6} presents the relation between the GTR centroid energy ratio $E^\text{cent.}_{K=1}/E^\text{cent.}_{K=0}$
and the axis ratio $a_0/b_0$.
Ratios extracted from microscopic calculations are shown as black spheres.
Red solid line represents the fitted function, and red shaded area indicates its $95\%$ confidence interval (CI).
A clear decreasing trend is observed in the centroid energy ratio with increasing axis ratio,
indicating that the deformation splitting in the GTR exhibits a geometric dependence similar to that in the IVGDR.
Such a behavior will be also understood in a microscopical way in Fig.~\ref{fig7}.

\paragraph*{Microscopic understanding of $K$-splitting in centorid energies and summed strengths in GTR regions.}
Fig.~\ref{fig7} presents a schematic illustration of
the $K=0$ and $K=1$ modes of GTR in $\beta^-$ direction for prolate (left) and oblate (right) cases.
In the spherical limit, GTRs are primarily governed by transitions from neutron $j_>$ spherical shells
to proton $j_<$ shells marked as black bars.
When deformation is introduced, these spherical orbitals split into deformed orbitals
with different projected angular momentum $\Omega$, shown as green (neutron) and red (proton) bars.
Solid bars denote states with $\Omega > 0$, while open bars represent their time-reversed states with $\Omega < 0$.
Transitions obeying the selection rules $\Omega_\text{final} - \Omega_\text{init.} = 0$
and $\Omega_\text{final} - \Omega_\text{init.} = 1$ are attributed to the $K=0$ and $K=1$ modes, respectively,
and illustrated by blue dashed and orange dotted arrows.

In $^{60\text{-}68}\text{Zn}$, the dominant transitions originate from neutron orbitals associated with the $1f_{7/2}$ shell,
which lie well below the Fermi surface and fully occupied.
For a given final state, as shown in panels (a) and (b) of Fig.~\ref{fig7},
the occupation probabilities of all initial states are equal ($v^2_\text{init.} = 1$),
and hence, occupation factors do not contribute to the centroid energy splitting between two modes.
Besides, the selection rules of two modes restricts the number of transitions are the same.
Consequently, only the single-particle energy difference between initial and final states can contribute to the centroid energy splitting.

To illustrate this, consider transitions to a final proton orbital with $\Omega^\text{final}_p = 1/2$,
which is degenerate with its time reversed state with $\Omega^\text{final}_{\bar{p}} = -1/2$.
For the $K=0$ mode, the selection rule $\Omega_\text{final} = \Omega_\text{init.}$ restricts
the initial neutron orbital to have the same $\Omega$ as the final state, i.e.,
$\Omega^\text{init.}_n = 1/2$ or $\Omega^\text{init.}_{\bar{n}} = -1/2$.
The single-particle transition energy  for $K=0$ is considered as a benchmark, and for simplification we set it to be zero,  as shown in Fig. ~\ref{fig7}.
In contrast, the $K=1$ mode allows two types of transitions:
(1) $\Omega_{\bar{n}} = -1/2 \rightarrow \Omega_p = 1/2$, and
(2) $\Omega_{\bar{n}} = -3/2 \rightarrow \Omega_{\bar{p}} = -1/2$.
Due to deformation, orbitals with larger $|\Omega|$ ($\Omega_{\bar{n}} = -3/2$)
are split to higher energies for prolate nuclei and lower energies for oblate nuclei.
As a result, the  $K=1$ transition involves a nonzero energy difference
corresponding to the gap between the $|\Omega| = 3/2$ and $|\Omega| = 1/2$ orbitals.
This yields a lower (higher) average excitation energy for the $K=1$ mode than that of the $K=0$ mode in prolate (oblate) nuclei.

For  final states with  arbitrary $\Omega_p \neq 1/2$, similar mechanism is found.
The $K=0$ mode always occurs in orbitals with identical $\Omega$, where the corresponding transition energy is set to zero as well.
On the other hand, the $K=1$ selection rule restricts,
$\Omega_n = \Omega_p - 1 \rightarrow \Omega_p$ and $\Omega_{\bar{n}} = \Omega_{\bar{p}} - 1 \rightarrow \Omega_{\bar{p}}$.
According to typical trends in the Nilsson diagram \cite{Firestone_1997_Table},
the energy splitting between $|\Omega|$ and $|\Omega + 1|$ is larger than that between $|\Omega|$ and $|\Omega - 1|$.
This also leads to the same systematic trend:
in prolate nuclei, the $K=1$ centroid lies below the $K=0$ centroid, while in oblate nuclei, the opposite is true.

In heavier isotopes, $^{72\text{-}78}\text{Zn}$, neutron orbitals from the $1g_{9/2}$ shell begin to contribute.
These transitions typically involve final proton states far above the Fermi surface and thus unoccupied ($v^2_\text{final.} = 0$).
As in the lighter isotopes, the effects from occupation probabilities and number of transitions are again negligible,
and the centroid energy splitting remains driven by single-particle level structures in deformed cases.
Similar analysis for understanding the relative single-particle transition energies between $K=1$ and $K=0$ modes
can be carried out but for given initial states with $\Omega_n = 1/2$ and $\Omega_n \neq 1/2$,  as shown in panels (c) and (d) of Fig.~\ref{fig7}.

In summary, the GTR centroid energy splitting of $K$-components arises from deformation splitting of single-particle levels with different $\Omega$ values.
The resonances predominantly involve either fully occupied or fully unoccupied states, ensuring that occupation probabilities play no role.
The selection rules ensure that the numbers of transitions in two modes are the same.
Thus, the splitting reflects an effect governed by the single-particle level structure.
This mechanism provides a microscopic explanation for systematic trends observed in centroid energies across the Zn isotopic chain,
as discussed in Figs.~\ref{fig5} and \ref{fig6}.

\begin{figure}
  \center
  \includegraphics[scale=0.8]{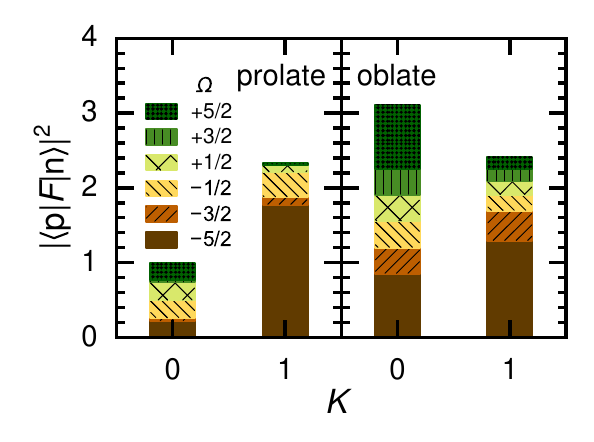}
  \caption{
    Sum of the squared transition matrix elements between neutron orbitals originating from the $1f7/2$ spherical shell
    and proton orbitals originating from the $1f5/2$ spherical shell in $^{66}\text{Zn}$.
    The left (right) panel corresponds to the prolate (oblate) case.
    Colored and filled patterns denote final states with different projected angular momentum ($\Omega$), and
    initial states are determined by the selection rule.
    \label{fig8}}
\end{figure}

After the microscopic illustration of the centroid energy splitting, in the following we try to understand
the behavior of $m_0$ for the $K=0$ and $K=1$ components in the resonance region.
As previously discussed in the context of centroid energies, the occupation probabilities and the number of selection-rule-allowed transitions
do not attribute to differences between the $K=0$ and $K=1$ modes.
This implies that the observed differences in $m_0$ must solely merge from differences between single-particle transition matrix elements of two modes.

Fig.~\ref{fig8} displays the squared transition matrix elements, $|\langle p|\hat{\Sigma}_K \hat{\tau}_-|n\rangle|^2$ associated with
initial neutron states originating from the $1f_{7/2}$ spherical shell and final proton states originating from the $1f_{5/2}$ shell in $^{66}\text{Zn}$.
These transitions contribute the majority of GTR strengths in $^{66}\text{Zn}$.
The left and right panels correspond to prolate and oblate configurations, respectively.
Each pattern in the column represents a transition to a proton final state with a specific $\Omega$.
The corresponding neutron initial states are determined by the selection rules.

One can observe a clear systematic:
in the prolate case, the summed transition strengths for the $K=0$ component are lower than
those for $K=1$, whereas in the oblate case, the opposite trend is seen.
This asymmetry is mainly driven by transitions involving final states with $|\Omega| = 5/2$.
For $K=0$ modes, the transitions are
$n: +5/2 \rightarrow p: +5/2$ and $n: -5/2 \rightarrow p: -5/2$;
For $K=1$ modes, the transitions are
$n: -7/2 \rightarrow p: -5/2$ and $n: +3/2 \rightarrow p: +5/2$.
Among these,
the contribution from the $n: +3/2 \rightarrow p: +5/2$ transition is found to be negligible,
so the comparison reduces to $n: \pm5/2 \rightarrow p: \pm5/2$ for $K=0$
and $n: -7/2 \rightarrow p: -5/2$ for $K=1$.

The key to understanding the change of transition strengths lies in
the spin mixing of the involved orbitals.
In the prolate configuration, the large components of
the neutron and proton wave functions $n:-5/2$ and $p:-5/2$
have nearly pure spin-down and spin-up orientations, respectively,
which results in a quite small transition matrix element for $K=0$ mode due to the poor spin overlap.
On the other hand, the large components of
the neutron and proton wave functions $n:-7/2$ and $p:-5/2$  also
have nearly pure spin-down and spin-up orientations, respectively,   which results in a large transition matrix element for $K=1$ operator. This explains the larger transition strength for $K=1$ mode compared to that for $K=0$ mode.
In contrast, in the oblate configuration,
the mixing of spin-up and spin-down components in the neutron and proton $|\Omega|=5/2$ wave functions
enhances the overlap of spin components and hence the transition matrix elements of $\langle p: \pm 5/2| F | n: \pm 5/2\rangle $ for $K=0$ mode.
However, this spin mixing also reduces the strength of the $K=1$ transition
from the nearly pure spin-down $n:-7/2$ orbital to the
final state $p:-5/2$ with less spin-up component.
As a consequence, the $m_0$ for the $K=0$ mode is smaller than that of the $K=1$ mode in prolate nuclei,
whereas the opposite trend is observed in oblate nuclei.
The deformation dependence of spin mixing provides a microscopic explanation for the observed trends in the summed strength of GTR.

\begin{figure}
  \center
  \includegraphics[scale=0.8]{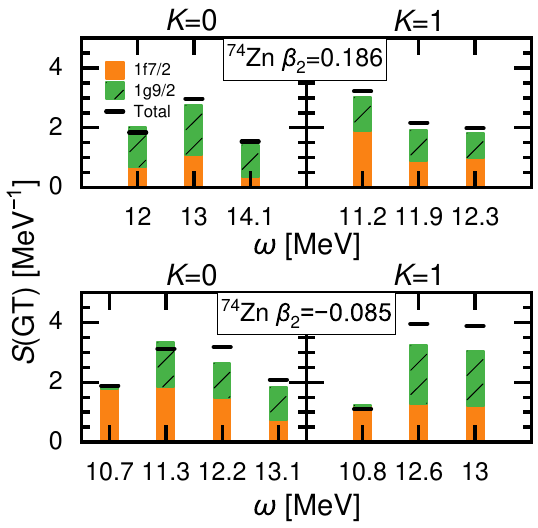}
  \caption{
    Decomposed strengths in terms of neutron initial states originating from
    different spherical shells in $^{74}\text{Zn}$.
    The upper (lower) panel displays the prolate (oblate) case.
    The black bars show the total strengths,
    and the orange (green) region represents strength contributed by
    initial neutron states from the $1f7/2$($1g9/2$) shell.
    \label{fig9}}
\end{figure}

In summary, the systematic trends of summed strengths and centroid energies in the GTR region between the $K=0$ and $K=1$ modes, 
as shown in Fig.~\ref{fig5}, were understood through single-particle transitions.
The opposite sign in centroid energy difference between $K=0$ and $K=1$ modes for prolate and oblate case is determined by the opposite ordering in the single-particle level splitting caused by deformation in prolate and oblate nuclei, as illustrated in Fig.~\ref{fig7}.
On the other hand, the summed strengths of these two modes are driven by the single-particle transition matrix elements (Fig.~\ref{fig8}), 
which are determined by the spin mixing of the involved orbitals.

\paragraph*{Microscopic understanding of the evolution of GTR peak strengths in $K=0$ and $K=1$ modes.}
Besides summed strengths and centroid energies,
the GTR peak strengths also exhibit a systematic trend in the lighter Zn isotopes $^{60\text{-}68}\text{Zn}$.
The $K=0$ peak strength is consistently lower than the $K=1$ peak strength in the prolate configuration,
but becomes higher in the oblate configuration, as shown in Fig.~\ref{fig4}.
However, this trend disappears in heavier isotopes $^{72\text{-}78}\text{Zn}$,
where the peak strengths of two modes become nearly comparable for opposite deformations.

To understand the disappearance of this trend in heavier isotopes, $^{74}\text{Zn}$ is examined as an example.
Fig.~\ref{fig9} displays the decomposed peak strengths,
highlighting contributions from neutron initial states originating from the $1f_{7/2}$
and $1g_{9/2}$ spherical shells for the prolate (upper panel) and oblate (lower panel) configurations.
The colored and filled patterns distinguish the spherical origins of the contributing initial neutron states.
It is observed that transitions associated with the $1f_{7/2}$ shell maintain the same behavior as in lighter isotopes.
The $K=0$ peak is lower (larger) in the prolate (oblate) case.
However, transitions originating from the $1g_{9/2}$ shell, which only occurs in heavier nuclei, exhibit the opposite trend.
These competing contributions cancel each another,
leading to nearly comparable $K=0$ and $K=1$ peak strengths in the resonance region of $^{72\text{-}78}\text{Zn}$.

\begin{figure*}
  \center
  \includegraphics[scale=0.8]{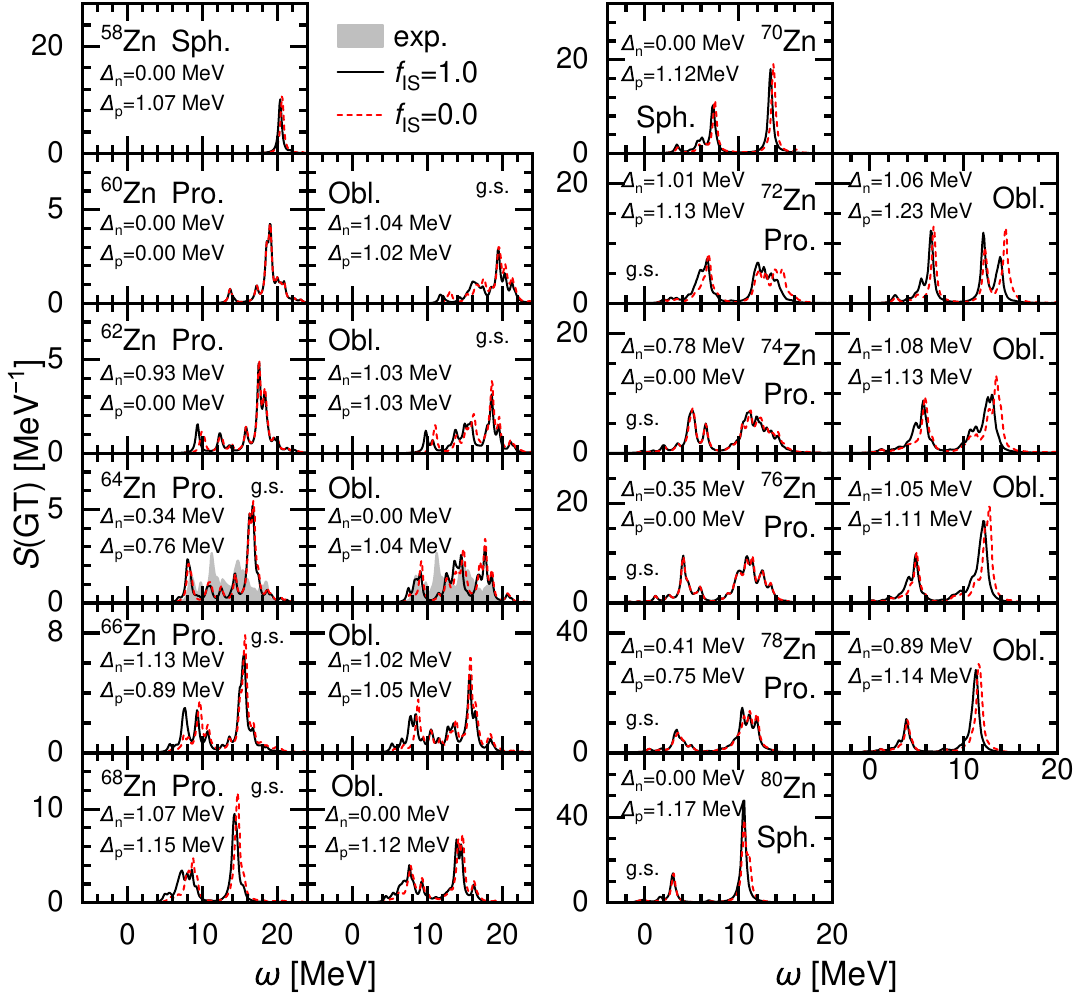}
  \caption{
    Comparison of GT strengths in Zn isotopes obtained with ($f_\text{IS}=1.0$)
    and without ($f_\text{IS}=0.0$) the isoscalar pairing interaction.
    The calculations are performed on top of both the ground states and local minima along the PECs.
    Results with isoscalar pairing are shown by black solid lines, while those without are shown by red dashed lines.
    For $^{64}\text{Zn}$, the experimental $B(\text{GT})$ distribution is smeared with a width of $0.25 \text{ MeV}$
    and scaled by a factor of $4.67$ (gray shaded area) according to the ratio of theoretical and experimental summed strengths \cite{FDiel_2019_PRC}.
    The average proton and neutron pairing gaps, $\Delta_p$ and $\Delta_n$, are also indicated,
    as defined in Ref.~\cite{Afanasjev_2015_PRC}.
    \label{fig10}}
\end{figure*}

\paragraph*{The isoscalar pairing effect on GT strengths.}
In the last part of discussions, the isoscalar pairing effects are investigated.
Fig.~\ref{fig10} shows the comparison of GT strengths in Zn isotopes
obtained with ($f_\text{IS}=1.0$) and without ($f_\text{IS}=0.0$) the isoscalar pairing interaction.
The calculations are performed on top of both the ground states and local minima along the PECs.
Results with isoscalar pairing are shown by black solid lines,
while those without are shown by red dashed lines.
For $^{64}\text{Zn}$ which has available experimental data,
the experimental $B(\text{GT})$ is smeared and scaled as in Fig.~\ref{fig4}
according to the ratio of theoretical and experimental summed strengths \cite{FDiel_2019_PRC}.
The average pairing gaps for protons and neutrons, $\Delta_p$ and $\Delta_n$, are also indicated,
following the definition given in Ref.~\cite{Afanasjev_2015_PRC}.

It can be seen that in the case of prolate $^{60}\text{Zn}$, prolate $^{64}\text{Zn}$, prolate $^{74-78}\text{Zn}$, 
which have vanishing or nearly vanishing neutron and proton pairing gaps,
the isoscalar pairing effect is negligible.
While for those with relatively large $\Delta$, the inclusion of isoscalar pairing shifts
the GT strength distribution toward lower excitation energies.
This shift may enhance the strengths located within the $Q_\beta$ window
and thus reduce the $\beta$-decay half-lives.

In lighter isotopes ($^{60\text{-}68}\text{Zn}$), except for $^{58}\text{Zn}$ where GT transitions occur only in the resonance region,
a systematic pattern emerges, that the isoscalar pairing tends to affect the low-lying GT strength more significantly,
whereas its influence in the resonance region is less pronounced.
This effect is quite pronounced in prolate $^{62}\text{Zn}$, where the resonance part remains unaffected.

In contrast, for heavier isotopes $^{70\text{-}80}\text{Zn}$,
the isoscalar pairing noticeably modifies the GT strength even in the resonance region.
This can be attributed to the angular momentum dependence of the isoscalar pairing matrix element.
In lighter isotopes, the resonance region is mainly dominated by transitions
involving neutron initial states from the $1f_{7/2}$ spherical shell ($l = 3$),
whereas in heavier isotopes, contributions from $1g_{9/2}$ orbitals ($l = 4$) become increasingly important.
Since the isoscalar pairing is more effective between states with higher orbital angular momentum,
its impact in the resonance region becomes more significant in $^{72\text{-}80}\text{Zn}$.

For the experimentally available $^{64}\text{Zn}$, in the absence of isoscalar pairing,
the GT strengths obtained with the oblate configuration are closer to the experimental distribution than the prolate case.
The isoscalar pairing acts on the oblate configuration, and shifts the low-lying peaks even closer to the experimental energies.
However, in the ground-state calculation, the prolate minimum yields the lowest total energy,
with an energy difference of $264\text{ keV}$ between the two minima.
The comparison of GT strengths with experimental data raises some doubts on the ground-state deformation from mean-field calculations.

\section{Conclusions\label{concl}}
In this work, the quasiparticle finite amplitude method (QFAM) is extended
to describe charge-exchange transitions based on relativistic Hartree-Bogoliubov model,
incorporating the relativistic point-coupling energy density functional DD-PC1
and a finite-range separable pairing force.

Deformation effects on Fermi and Gamow-Teller (GT) transitions in Zn isotopes have been systematically studied.
The potential energy curves show significant shape evolution,
and the corresponding GT strength distributions reveal clear deformation splitting of the $K=0$ and $K=1$ components
and show obviously different strength distributions between prolate and oblate configurations.

The analysis of non-energy-weighted and energy-weighted summed strengths in GTR region ($m_0$ and $m_1$)
further shows the impact of nuclear deformation on GTR strength,
with prolate shapes favoring stronger $K=1$ contributions and oblate shapes enhancing $K=0$ strengths.
Moreover, the centroid energies of two components are also split, with the favored modes locating at lower energies.
The systematic trends in GTR peak strengths also reveal a distinct deformation dependence.
In lighter isotopes where transitions involving $\nu 1f_{7/2}$ spherical shell dominate, the $K=0$ peak strength is lower than the $K=1$ one in prolate configurations
but becomes higher in oblate configurations; while this trend disappears in heavier isotopes with the appearance of more transitions involving $\nu 1g_{9/2}$ spherical shell.
Above systematic behaviors have been understood via the single-particle level structure in the deformed nuclei.

Finally, the isoscalar pairing effect on the GT strength has been examined.
The isoscalar pairing strength is found to shift the GT strength toward lower energies, and
this shift becomes more pronounced in heavier isotopes where the higher-$l$ transitions are involved.

\begin{acknowledgments}
  The author (C.C.) would like to thank Dr.~Ante Ravli\'{c} for fruitful discussions on numerical tests.
  This work was supported by
  the National Key Research and Development Program under Grant No.~2021YFA1601500,
  the Lingchuang Research Project of China National Nuclear Corporation under Grant No.~CNNC-LCKY-2024-082,
  the National Natural Science Foundation of China under Grants No.~12075104, No.~12447106, No.~12305129, and No.~12405132
  and the Fundamental Research Funds for the Central Universities under Grant No.~lzujbky-2023-stlt01.
\end{acknowledgments}

\end{document}